
\input phyzzx
\hoffset=0.2truein
\voffset=0.1truein
\hsize=6truein
\def\TITLEPAGE{\frontpagetrue}
\def\CALT#1{\hbox to\hsize{\tenpoint \baselineskip=12pt
        \hfil\vtop{
        \hbox{\strut CALT-68-#1}}}}

\def\CALTECH{
        \address{California Institute of Technology,
Pasadena, CA 91125}}

\def\AUTHOR#1{\vskip .2in \centerline{#1}}

\def\ABSTRACT#1{\vskip .2in \vfil \centerline{\twelvepoint
\bf Abstract}
        #1 \vfil}
\def\ENDTITLEPAGE{\vfil\eject\pageno=1}

\tolerance=10000
\hfuzz=5pt

\def\Dslash{/\!\!\!\!D}

\def\vslash{\rlap{/}v}
\def\>{{\buildrel >\over {{}_{\sim}}}}
\def\<{{\buildrel <\over {{}_{\sim}}}}
\def\-{{\buildrel <\over {{}_{-}}}}
\def\Q{{(Q)}}
\TITLEPAGE
\CALT{1901}         
\bigskip
\titlestyle {Heavy Flavor Theory: Overview\foot{Work supported in part by the
U.S. Dept. of Energy
under Grant no. DE-FG03-92-ER 40701.}}
\AUTHOR{Mark B. Wise}
\CALTECH
\bigskip
\centerline{Talk given at the Lepton Photon Conference}
\centerline{Cornell University, August 1993}

\ABSTRACT{
An introduction to the heavy quark effective theory and its symmetries
is given.  Some implications of the heavy quark spin and flavor symmetries
are discussed.  Recent results on fragmentation to quarkonium states are
reviewed.}
\ENDTITLEPAGE
\eject

\noindent {\bf 1.  Introduction}

In the last few years there has been considerable progress in understanding
the physical properties of hadrons containing a single heavy quark. Much of
this progress has arisen from the development of the heavy quark effective
theory (HQET) and the application of its spin-flavor
symmetries.$^{1,2}$  The heavy
quark symmetries have implications for the physics of hadrons containing a
single heavy quark in a kinematic regime where nonperturbative strong
interaction physics is important. Here the situation is analogous to the
application of the approximate light quark flavor symmetries (e.g., isospin or
chiral $SU(2)_L\times SU(2)_R$). In this talk I will review the basic
elements of the heavy quark effective theory and give some applications of
its symmetries.

Like isospin or $SU(3)$ symmetry the heavy quark symmetries are approximate.
Of the six quarks that exist in nature the top, bottom and charm can
potentially be treated as heavy. The heavy quark symmetries arise in the
limit of QCD where the heavy quark masses $m_Q$ are taken to infinity. In the
real world, where the heavy quark masses are finite, corrections to the
predictions based on heavy quark symmetry of order $\Lambda_{QCD}/m_Q$ arise.
Here $\Lambda_{QCD}$ refers to a typical hadronic scale not necessarily the
parameter that occurs in the strong coupling constant. The top quark is very
heavy, however, it's lifetime is so short that ideas based on heavy quark
symmetry have little relevance for it's properties.  (It doesn't live long
enough to form a hadron.) The charm quark mass is not large enough for us to
have confidence that the $\Lambda_{QCD}/m_c$ corrections are negligible.
However,
by comparing the predictions of heavy quark symmetry with experiment
eventually  it will be possible to determine how good the $m_c
\rightarrow \infty$ limit is.  There are already several indications
that heavy quark spin symmetry is a useful concept for the charm quark,
however, the applicability of the flavor symmetry that relates the
properties of hadrons containing a charm quark to those containing a
bottom quark has hardly been tested.

In the next section two derivations of the heavy quark effective theory are
presented. One derives the HQET by taking a limit of the Feynman rules of QCD
to get the Feynman rules of the effective theory. The second derivation
relates the heavy quark fields in the full theory to those of the effective
theory. Spectroscopic applications are presented in the third section. This
includes a discussion of some very recent work on the application of
heavy quark symmetry to fragmentation. The fourth section contains a discussion
of
the sources of symmetry breaking. Applications to matrix elements relevant
for exclusive weak semileptonic decays are discussed in Section 5. This
includes a brief
presentation of some predictions that arise when chiral $SU(2)_L\times
SU(2)_R$ symmetry is combined with heavy quark symmetry. In Section 6 some
very recent work on the application of HQET to inclusive semileptonic
$B$ decays is discussed. Unfortunately it indicates that a model independent
extraction of $V_{ub}$ (the $b\rightarrow u$ Cabibbo-Kobayashi-Maskawa
matrix element) from the endpoint region of the electron energy spectrum
in semileptonic $B$ decays is not possible.

Not all the recent exciting developments in heavy flavor theory are
related to the heavy quark effective theory.  There have also been
recent advances in our understanding of the properties of $\bar Q Q$
quarkonium states.  For example, over the  past year it was realized
that fragmentation functions like $D_{c \rightarrow \psi} (z)$ are
computable in terms of  the $\psi$ wave function at the origin and the
charm quark mass (for essentially the same reason that $\psi$ decay to
three gluons is computable).  Fragmentation to $Q \bar Q$ states is
discussed in the final section.

\noindent{\bf 2.  The Heavy Quark Effective Theory}

Consider the situation where a heavy quark $Q$ (i.e., $m_Q
\gg\Lambda_{QCD}$) is interacting with light
degrees of freedom associated with a momentum scale much less than the heavy
quark mass.  Then it is appropriate to take the limit of QCD where
$m_Q\rightarrow
\infty$ with the heavy quark's four velocity $v^{\mu}$ held
fixed.$^{1,2}$  This
kinematic situation does occur in nature.  For example in its rest frame
 a $B^-$ meson has
the $b$-quark almost at rest at the center of the meson.  The size
of the meson, however, is determined by nonperturbative strong interactions and
is of
order $1/\Lambda_{QCD}$.  (For example, in a simple string picture it is the
tension of the QCD string that goes between the $b$-quark and the
$\bar u$-quark that fixes the size of the meson.)  Hence by the uncertainty
principle the typical momentum of the light degrees of freedom is of order
$\Lambda_{QCD}$.

One way of deriving the effective theory that results from this limit is to
directly take the limit of the Feynman rules of QCD. To do this we write the
heavy quark's four momentum as
$$ p_Q = m_Qv + k ~,\eqno (1)$$
$k$ is called
the residual momentum and is a measure of how much the heavy quark is
off-shell. The QCD propagator for the heavy quark is
$${i(\rlap{/}p_Q + m_Q)\over (p_Q^2-m_Q^2 + i\varepsilon)} ~. \eqno (2)$$
Now substitute eq. (1) into this
neglecting the residual momentum $k$ (in comparison with $m_Q$) wherever
possible. In the numerator of
eq. (2) the residual momentum can be neglected giving $m_Q(\rlap{/}v + 1)$.
However, if the residual momentum is neglected in the denominator we get
zero. Thus the leading part that survives in the denominator is the
piece that is linear in the residual momentum $2m_Qvk$. Putting our
expressions for the numerator and denominator together
we get the propagator for the effective  heavy quark theory
$${i(\rlap{/}v + 1)\over 2(v \cdot k + i\varepsilon)} ~. \eqno(3)$$
Note that it is independent of the heavy quark mass.  The vertex for
gluon heavy quark interactions is
$$- ig_s T^A \gamma_\mu \,\, , \eqno (4)$$
in QCD.  This vertex always appears between propagators in the
calculation of any Green function.  Because of (3) we can, in the
effective theory where $m_Q \rightarrow \infty$, replace (4) by
$$- ig_s T^A {(\vslash + 1)\over 2} \gamma_\mu {(\vslash + 1)\over 2} \,\, .
\eqno (5)$$
Anticommuting the $\vslash$ on the right through the $\gamma_\mu$ using
$\gamma^\mu (\vslash + 1) = 2v^\mu - (\vslash - 1) \gamma^\mu$, eq.
(5) becomes
$$-ig_s T^A v_\mu \,\, . \eqno (6)$$

The propagator in eq. (3) only has a single pole in the complex $k^0$
plane.  Thus heavy quark loops vanish since the contour of the $k^0$
integral can always be closed in the upper half plane giving zero.
There is no heavy quark pair creation in the effective theory.

With the vertex for gluon interactions of the form in eq. (6) factors
of the projection operator $(\vslash + 1)/2$ in the heavy quark
propagator can be moved to the outside of any Feynman diagram.  Then the
heavy quark propagator becomes
$${i\over v \cdot k + i\varepsilon} \,\, . \eqno (7)$$

Eqs. (6) and (7) are the Feynman rules for the HQET.  They are
independent of the heavy quark mass $m_Q$.  Note also that the gamma
matrices have completely disappeared indicating that the heavy quark's
spin is conserved.

It is also instructive to understand the relation between heavy quark
fields in the HQET and fields in QCD.  For QCD the part of the Lagrange
density that contains the heavy quark field $Q$ is
$${\cal L} = \bar Q (i\Dslash - m_Q)Q \eqno (8)$$
where
$$D_\mu = \partial_\mu + ig_s T^A A_\mu^A \,\, , \eqno (9)$$
is the covariant derivative.  Writing$^{2}$
$$Q = e^{-im_Q v \cdot x} h_v^{(Q)}\,\, , \eqno (10)$$
where
$$\vslash h_v^{(Q)} = h_v^{(Q)}\,\, , \eqno (11)$$
The Lagrange density in eq. (8) becomes
$${\cal L} =\bar h_v^{(Q)} (m_Q (\vslash - 1) + i\Dslash) h_v^{(Q)}$$
$$= \bar h_v^{(Q)} i\Dslash h_v^{(Q)}\,\, .~~~~~~~~~~~~~~~ \eqno (12)$$
Finally using the constraint in eq. (11) to insert factors of
$(\vslash + 1)/2$ on either side of the $\Dslash$ in the last line of eq.
(12) the Lagrange density becomes$^{2}$
$${\cal L} = \bar h_v^{(Q)} i v \cdot D h_v^{(Q)} \,\, . \eqno (13)$$
The Lagrange density in eq. (13) reproduces the Feynman rules in eqs.
(6) and (7) and eqs. (10) and (11) give the relation between the
heavy quark field in the effective theory, $h_v^{(Q)}$, and the heavy
quark field in the full theory $Q$.

Strong interactions of a heavy quark depend on its four velocity $v_\mu$
but not on its spin or mass.  Hence, if there are $N_f$ heavy quarks with
the same four velocity, then the effective heavy quark theory has a
SU(2N$_f$) spin flavor symmetry.  In practice it is the charm and bottom
quarks for which this symmetry is relevant.  Because $m_c = 1.5 GeV$ is not
that large we expect significant $\Lambda_{QCD}/m_c$ corrections to the
predictions of heavy quark symmetry (which become exact in the $m_c
\rightarrow \infty$ limit).

In the following sections several applications of heavy quark symmetry
are given.  Corrections to the predictions of heavy quark symmetry are
also discussed.  Applications of heavy quark symmetry fall into two
broad classes, spectroscopic applications and applications to weak decays.

\noindent{\bf 3.  Spectroscopy}

Because of heavy quark spin symmetry (in the rest frame of the heavy
quark) both the spin of the heavy quark $\vec S_Q$ and the total angular
momentum $\vec S$ commute with the Hamiltonian.  For the spectroscopy of
hadrons containing a single heavy quark $Q$ the spin of the light
degrees of freedom
$$	\vec S_\ell = \vec S - \vec S_Q\,\, , \eqno (14)$$
plays an important role.  Since both $\vec S$ and $\vec S_Q$ commute with the
Hamiltonian $\vec S_\ell$ also commutes with the Hamiltonian.  Thus
states at rest are labeled not only by their total spin $s$ but also by
the spin of the light degrees of freedom $s_\ell$.$^{[3]}$  For example the
ground state heavy mesons with $Q{\bar q} (q = u ~{\rm or}~ d)$ flavor
quantum numbers have $s_\ell = 1/2$ and negative parity.  Combining this
spin of the light degrees of freedom with the spin of the heavy quark
$s_Q = 1/2$ gives meson states $|P^{(Q)}\rangle$ and $|P^{*(Q)}\rangle$
that have total spins $s=0$ and $s = 1$ respectively.  For $Q = c$ these
are the $D$ and $D^*$ mesons while for $Q = b$ they are the $B$ and
$B^*$ mesons.  Writing
$$	|P^{(Q)}\rangle = {1\over\sqrt{2}} \left\{|\uparrow\rangle_Q
|\downarrow \rangle_\ell - |\downarrow\rangle_Q |
\uparrow\rangle_\ell\right\} \eqno (15a)$$
$$	|P^{*(Q)}\rangle = {1\over \sqrt{2}} \left\{|\uparrow\rangle_Q |
\downarrow\rangle_\ell + |\downarrow \rangle_Q |\uparrow\rangle_\ell\right\}
\eqno (15b)$$
it is easy to see that
$$	S_Q^3 |P^{(Q)}\rangle = {1\over 2} |P^{*(Q)}\rangle\,\, . \eqno
(16)$$
In eqs. (15) the first arrow refers to the spin of the heavy quark
along the quantization axis $(\hat z)$ and the second arrow refers to the spin
of
the light degrees of freedom along the quantization axis.  In eqs.
(15b) and (16) $|P^{*(Q)}\rangle$ refers to the spin one state with
zero spin along the quantization axis.  Eq. (16) implies that the
$|P^{(Q)}\rangle$ and $|P^{*(Q)}\rangle$ states have the same mass.

In general the spectrum of hadrons containing a single heavy quark $Q$
has (in the $m_Q \rightarrow \infty$ limit) for each spin of the light
degrees of freedom $s_\ell$ a degenerate doublet with total spins $s_+$
and $s_-$ where
$$	s_\pm = s_\ell \pm 1/2\,\, , \eqno (17)$$
(Except for the case $s_\ell = 0$ where the total spin must be $1/2$.)

States in a doublet associated with $s_\ell$ can decay to states in a
lower doublet associated with $s'_\ell$ by emission of light quanta
(e.g., $\pi, \pi\pi, \eta, \eta', \rho$, etc.) with total angular momentum
$L$.  The decay amplitudes for the possible transitions are related
by$^{3,4}$
$$	{\cal A} (s, s^3 \rightarrow s' s^{\prime 3} + L m)$$
$$	= R_{s_{\ell}, s'_{\ell}, L} (-1)^s \sqrt{2s_\ell + 1} \sqrt{2s'
+1} \left\{\matrix{s'_\ell & s_\ell & L\cr s & s' & 1/2\cr}\right\}$$
$$	\cdot \langle L, m;s', s^{\prime 3}|s, s^3\rangle\,\, , \eqno (18)$$
in the $m_Q \rightarrow \infty$ limit.  Here $R$ is a reduced matrix
element that depends on $s_\ell, s'_\ell, L$ and other possible quantum
numbers that label the doublets.  It is not surprising that the four
transitions $s_\pm = s'_\pm + Lm$ are determined by a single reduced
matrix element.  Because of heavy quark spin symmetry there is really
only one fundamental transition  $s_\ell \rightarrow s'_\ell + Lm$.  The
angular distribution of the decay products is determined by
$$	{\cal A} (\Omega) = \sum_m Y_{Lm} (\Omega) {\cal A} (s,s^3 \rightarrow
s', s^{\prime 3} + Lm) \,\, , \eqno (19)$$
for fixed $s^3$ and $s^{\prime 3}$.

For $Q = c$ an excited heavy meson multiplet with $s_\ell = 3/2$ and
positive parity has been observed.  It is composed of the spin two state
$D_2^* (2460)$ and the spin one state $D_1 (2420)$.  Eq. (18) predicts
that the relative partial widths
$$ \Gamma (D_2^* \rightarrow [\pi D]_{L = 2}): ~\Gamma (D_2^* \rightarrow
[\pi D^*]_{L = 2}):  ~\Gamma (D_1 \rightarrow [\pi D^*]_{L = 2}) \,\, ,
\eqno (20)$$
are
$$	2/5 \qquad \qquad : \qquad \qquad 3/5 \qquad \qquad : \qquad 1
\eqno (21)$$
and that
$$	\Gamma (D_1 \rightarrow [\pi D^*]_{L = 0}) = 0 \,\, . \eqno (22)$$

Eqs. (20), (21) and (22) hold in the $m_c \rightarrow \infty$ limit.
There is, however, an important $\Lambda_{QCD}/m_c$ correction of
kinematic origin that must be taken into account.  For small $|\vec
p_\pi|$ we expect the momentum dependence of these decay amplitudes to
be of the form $|\vec p_\pi|^{2L + 1}$.  Since the $D^* - D$ mass
difference is not very small compared with the $D - D^*_2$ mass difference
and  the power $2L + 1 = 5 $ (for $L = 2$) is large it is a bad
approximation to treat $|\vec p_\pi |$ as the same for the four
transitions.  Thus, when comparing with experiment, we must multiply eq.
(21) by factors of $|\vec p_\pi|^5$.  These factors are
$$	3.3 \qquad \qquad : \qquad \qquad 0.88 \qquad \qquad : \qquad\qquad
0.57 \,\, \eqno (23)$$
in units of $10^{-2} GeV^5$.  Multiplying (21) by (23) gives
predictions for the relative decay rates that include the spin symmetry
violation arising from the $D^* - D$ and $D_2^* - D_1$, mass differences.
For example,
$$	\Gamma (D_2^* \rightarrow D \pi)/\Gamma (D_2^* \rightarrow D^*
\pi) = 2.5 \eqno (24)$$
follows from eqs. (21) and (23) and is in good agreement with the
experimental value $2.4 \pm 0.7$.  Eq. (22) is also in agreement with
observation.  If the $S$-wave $D_1 \rightarrow \pi D^*$ amplitude were
the ``typical'' size expected on the basis of our experience with
hadronic physics, the $D_1$ width would be
${\buildrel >\over {{}_{\sim}}}$ 100 MeV.  Since the
$D_1$ width is only about 20 MeV the $S$-wave $D_1 \rightarrow \pi D^*$
amplitude is small.$^{5}$  However because $D$-wave decay amplitudes are
usually much smaller than $S$-wave decay amplitudes it is possible that
the small order $\Lambda_{QCD}/m_c$ $S$-wave $D_1 \rightarrow \pi D^*$
amplitude contributes significantly to the $D_1$ width.

Heavy quark spin symmetry relates fragmentation probabilities for
members of a doublet.  It implies that the probability,
$P_{Q,h_{Q}\rightarrow s,h_{s}}^{(H)}$, of a heavy quark $Q$ with helicity
$h_Q$ (along the fragmentation axis) fragmenting to heavy
hadron $H$ in a doublet with spin of the light degrees $s_\ell$, total
spin $s$ and total helicity (along the fragmentation axis) $h_s$
is$^{4}$
$$	P_{Q,h_{Q} \rightarrow s,h_{s}}^{(H)} = P_{Q \rightarrow
s_{\ell}} \sum_{h_{\ell}} p_{h_{\ell}} |\langle s_Q, h_Q; s_\ell, h_\ell
| s, h_s \rangle |^2 \eqno (25)$$
Here $P_{Q \rightarrow s_{\ell}}$ is the probability of the heavy quark
fragmenting to a doublet with spin of the light degrees of freedom
$s_\ell$.  It is independent of the heavy quarks helicity but will
depend on other quantum numbers needed to specify the doublet.  $p_\ell$
is the conditional probability that the light degrees of freedom have
helicity $h_\ell$ (given that $Q$ fragments to $s_\ell$).  The probability
interpretation implies that $ 0 \- p_{h_{\ell}} \- 1$ and
$$	\sum_{h_{\ell}} p_{h_{\ell}} = 1 \,\, . \eqno (26)$$

Parity invariance of the strong interactions implies that
$$	p_{h_{\ell}} = p_{-h_{\ell}}\,\, . \eqno (27)$$
Eqs. (26) and (27) restrict the number of independent probabilities
$p_{h_{\ell}}$ to be equal to $s_\ell - 1/2$ for mesons and $s_\ell$ for
baryons.

Parity invariance of the strong interactions implies that
$$	P_{Q, h_{Q} \rightarrow s, h_s}^{(H)} = P_{Q, - h_{Q}
\rightarrow s, - h_{s}}^{(H)}\,\, . \eqno (28)$$
Heavy quark spin symmetry reduces the number of independent
fragmentation probabilities.  For mesons with spin of the light degrees
of freedom $s_\ell$ the fragmentation probabilities, $P_{Q, h_{Q}
\rightarrow s, h_{s}}^{(H)}$, are expressed in terms the $s_\ell - 1/2$
($s_\ell$ for baryons) independent $p_{h_{\ell}}$'s and $P_{Q \rightarrow
s_{\ell}}$.

For the $D$ and $D^*$ mesons $s_\ell = 1/2$ and eqs. (26) and (27)
give $p_{1/2} = p_{- 1/2} = 1/2$.  The relative fragmentation probabilities
$$	P_{c, 1/2 \rightarrow 0,0}^{(D)} ~~:~~ P_{c, 1/2 \rightarrow
1,1}^{(D^*)} ~~:~~ P_{c, 1/2 \rightarrow 1,0}^{(D^*)} ~~:~~ P_{c, 1/2
\rightarrow 1,-1}^{(D^*)} \,\, \eqno (29)$$
are
$$	1/4 \qquad ~~:~~~~ 1/2 \qquad ~~:~~~~ 1/4 \qquad ~~:~~~~ 0 \,\, .~\qquad
\eqno (30)$$
Eq. (28) determines the fragmentation probabilities for a helicity
$-1/2$ charm quark in terms of those above.  The relative fragmentation
probabilities for the three $D^*$ helicities agree with experiment,
however, the prediction that the probability for a charm quark to
fragment to a $D$ is one third the probability to fragment to a $D^*$
does not.  This violation of  heavy quark spin symmetry may have its
origin in the $D^* -D$ mass difference which suppresses the
fragmentation to $D^*$'s.

Fragmentation probabilities for hadrons containing different heavy
quarks $Q$ are equal by heavy quark flavor symmetry.  In the $m_Q
\rightarrow \infty$ limit the shape of a heavy hadrons fragmentation
function is, $\delta (1 - z)$, since the heavy quark carries all the
hadron's
momentum (this is for the fragmentation function $D_{Q \rightarrow H}
(z)$ evaluated at a subtraction point $\mu \approx m_Q$).  For finite
$m_Q$ the fragmentation functions support is concentrated in a region of
$z$  near one of order $\Lambda_{QCD}/m_Q$.$^{6}$

\noindent{\bf 4.  $\Lambda_{QCD}/m_Q$ Corrections}

The sources of heavy quark symmetry breaking that arise because of the
finite value of $m_Q$ are understood.  Including effects of order
$\Lambda_{QCD}/m_Q$ the Lagrange density for the heavy quark effective
theory is$^{7}$
$${\cal L} = \bar h_v^{(Q)} iv \cdot D h_v^{(Q)} + {1\over 2m_Q} \bar
h_v^{(Q)} (iD)^2 h_v^{(Q)}$$
$$	-a_2 {1\over 4m_Q} \bar h_v^{(Q)} g_s \sigma^{\mu\nu}
G_{\mu\nu}^A T^A h_v^{(Q)} + {\cal O} (1/m_Q^2) \,\, . \eqno (31)$$
In the leading logarithmic approximation$^{8}$
$$	a_2 (\mu) = \left[{\alpha_s (m_Q)\over \alpha_s
(\mu)}\right]^{9/(33-2N_f)}\,\, . \eqno (32)$$
The corrections to the $m_Q \rightarrow \infty$ limit have a simple
physical origin.  The second term in eq. (31) is the kinetic energy of
the heavy quark.  It breaks the heavy quark flavor symmetry but not the
spin symmetry.  The last term is the energy that arises from the
chromomagnetic moment of the heavy quark.  The factor $a_2(\mu)$ arises
because the operator $\bar h_v^{(Q)} g_s \sigma^{\mu\nu} G_{\mu\nu}^A
T^A h_v^{(Q)}$ requires renormalization.  There is no renormalization
point dependence to the heavy quark kinetic energy coefficient because
of reparametrization invariance.$^{9}$  This last term breaks both the heavy
quark spin and flavor symmetries.

The second and third terms in eq. (31) influence hadronic masses. It is
convenient to introduce dimensionless quantities $K_Q(H^{(Q)})$ and $G_Q
(H^{(Q)})$ that take this into account.  For hadronic states
$|H^{(Q)}\rangle$ normalized to unity we define
$$	K_Q (H^{(Q)}) = {1\over m_Q^2} \langle H^{(Q)}| {1\over 2} \bar
h_v^{(Q)} D^2 h_v^{(Q)}| H^{(Q)}\rangle \eqno (33a)$$
and
$$	G_Q (H^\Q) = {a_2\over m_Q^2} \langle H^\Q| {1\over 4} \bar
h_v^\Q g_s \sigma^{\mu\nu} G_{\mu\nu}^A T^A h_v^\Q| H^\Q\rangle\,\, .
\eqno (33b)$$
$K_Q$ is proportional to $(\vec D h_v)^{\dag} (\vec D h_v)$ and doesn't
(because of reparametrization invariance) have any subtraction in its
definition.  It follows that $K_Q$ is positive.

The matrix elements in (33) are evaluated in the $m_Q
\rightarrow \infty$ limit and are independent of the heavy quark
masses.  It is the coefficients $1/m_Q^2$ and $a_2(\mu)/m_Q^2$ that
contain the dependence on $m_Q$.

The $s_\ell = 1/2$ negative parity ground state meson doublet has spin
zero and spin one mesons which we denoted by $P^\Q$ and $P^{*(Q)}$.  Up
to terms of  order $(\Lambda_{QCD}/m_Q)^2$ the masses of these mesons
are
$$	M(P^\Q) = m_Q + \bar\Lambda (P^\Q) + m_Q K_Q (P^\Q) + m_Q G_Q
(P^\Q)\eqno (34a)$$
$$	M(P^{*(Q)}) = m_Q + \bar\Lambda (P^\Q) + m_Q K_Q (P^\Q) -
{1\over 3} m_Q G_Q (P^\Q) \,\, . \eqno (34b)$$
The lowest lying positive parity $s_\ell = 3/2$ excited meson doublet
has spin two and spin one members that I denote by $P_2^{*(Q)}$ and
$P_1^\Q$ (for $Q=c$ there are the $D_2^* (2460)$ and $D_1(2420)$ mesons).
The masses of these hadrons are$^{10}$
$$	M(P_1^\Q) = m_Q + \bar\Lambda (P_1^\Q) + m_Q K_Q (P_1^\Q) + m_Q
G_Q (P_1^\Q) \eqno (35a)$$
$$	M(P_2^{*(Q)}) = m_Q + \bar\Lambda (P_1^\Q) + m_Q K_Q (P_1^\Q) -
{3\over 5} m_Q G_Q (P_1^\Q) \,\, . \eqno (35b)$$
In eqs. (34) and (35) $\bar \Lambda$ is a positive contribution to the heavy
meson mass that is independent of $m_Q$.  It comes from the part of the
QCD Hamiltonian involving only the light degrees of freedom.  A rigorous
lower bound on $\bar\Lambda (P^\Q)$ has recently been derived.$^{11}$   In
eqs. (34) and (35) we used $G_Q (P^\Q) = - 3 G_Q (P^{*(Q)})$ and $G_Q
(P_1^\Q) = - (5/3) G_Q (P_2^{*(Q)})$.  These relations follow from the
fact that hadronic matrix elements of $\bar h_v^\Q g_s \sigma^{\mu\nu}
G_{\mu\nu}^A T^A h_v^\Q$ are proportional to
$$	2\vec S_\ell \cdot \vec S_Q = [\vec S^2 - \vec S_\ell^2 - \vec S_Q^2]
= [s(s+1) - s_\ell (s_\ell + 1) - 3/4]\,\, . $$
This operator causes the splitting between members of doublets
$$	m_Q G_Q (P^\Q) = {3\over 4} [M(P^\Q) - M(P^{*(Q)})] \eqno (36a)$$
$$	m_Q G_Q (P_1^\Q) = {5\over 8} [M(P_1^\Q) - M(P_2^{*(Q)})] \,\, .
\eqno (36b)$$
Comparing eqs. (36) with the measured $D, D^*, D_1$ and $D_2^*$ masses
gives $G_c (D) \simeq -0.06$ and $G_c(D_1) = - 0.02$.  The fact that
these order ($\Lambda_{QCD}/m_c)^2$ quantities are small indicates that
the $m_c \rightarrow \infty$ limit is a useful approximation.  From the
known dependence of $G_Q (P^\Q)$  on the heavy quark mass it follows
that$^{8}$
$$	(m_{B^{*}} - m_B) = \left({m_c\over m_b}\right) \left[{\alpha_s
(m_b)\over \alpha_s (m_c)}\right]^{9/25} (m_{D^{*}} - m_D) \,\, ,\eqno
(37)$$
which is in agreement with experiment.  It is probably a
mistake to view this success as very important since a similar formula is
known to hold even for light hadrons.  However, the validity of eq. (37)
does provide
some support for the usefulness of heavy quark flavor symmetry in
relating properties of hadrons containing a $b$-quark to those
containing a $c$-quark.  We also
have
$$	(m_{B_{2}^{*}} - m_{B_{1}}) = \left({m_c\over m_b}\right)
\left[{\alpha_s (m_b)\over \alpha_s (m_c)}\right]^{9/25} (m_{D_{2}^{*}} -
m_{D_{1}})\,\, . \eqno (38)$$
We can take a linear combination of eqs. (34) and of eqs. (35) for
which $G_b$ cancels out.  Define
$$	M(P^\Q)_{avg} = {M(P^\Q) + 3M(P^{*(Q)})\over 4} \eqno (39a)$$
$$	M(P_1^\Q)_{avg} = {3M(P_1^\Q) + 5 M(P_2^{*(Q)})\over 8} \eqno
(39b)$$
$$	M(P^\Q)_{avg} = m_Q + \bar\Lambda (P^\Q) + m_Q K_Q (P^\Q) \eqno
(40)$$
$$	M(P_1^\Q)_{avg} = m_Q + \bar\Lambda (P_1^\Q) + m_Q K_Q
(P_1^\Q)\,\, . \eqno (41)$$
These relations (and the known dependence of $K_Q$ on $m_Q$) imply, for
example that,
$$	[M(D_1)_{avg} - M(D)_{avg}] - [M(B_1)_{avg} - M(B)_{avg}]$$
$$	= m_b \left({m_b\over m_c} -1\right) (K_b (B_1) - K_b (B)) \,\,
. \eqno (42)$$

Similar mass formulae hold for the baryons.  They are particularly simple
for the ground state isospin zero baryon which has $s_\ell = 0$.  For
this $\Lambda_Q$ baryon $G_Q (\Lambda_Q) = 0$ and
$$	M(\Lambda_Q) = m_Q + \bar\Lambda (\Lambda_Q) + m_Q K_Q
(\Lambda_Q)\,\, . \eqno (43)$$
Combining eqs. (43), (39a) and (34) gives
$$	[M(\Lambda_c) - M(D)_{avg}] - [M(\Lambda_b) - m(B)_{avg}]$$
$$	= m_b \left({m_b\over m_c} -1\right) (K_b (\Lambda_b) - K_b (B))
\,\, . \eqno (44)$$
The measured values of the $\Lambda_c, \Lambda_b, D, D^*, B$ and $B^*$
masses implies from eq. (44) that the difference between the heavy
quark kinetic energies in the $\Lambda_b$ and $B$ is small.

\noindent{\bf 5.  Exclusive Semileptonic Decays}

Heavy quark flavor symmetry plus isospin symmetry implies (again heavy
meson states are normalized to unity) that
$$	\langle L(k,\epsilon) | \bar u \gamma_\mu (1 - \gamma_5) b| B(v)
\rangle$$
$$	= \left[{\alpha_s (m_b)\over \alpha_s (m_c)}\right]^{-6/25}
\langle L (k,\epsilon) | \bar d \gamma_\mu (1 - \gamma_5) c|
D(v)\rangle\,\, \eqno (45)$$
where $L$ is a state that doesn't contain a heavy quark (i.e., $L$ =
vacuum, pion, rho, etc.).  The factor of $[\alpha_s (m_b)/\alpha_s
(m_c)]^{-6/25}$ in eq. (45) arises from the relationship between
currents in the full theory of QCD and currents in the effective
theory$^{12}$
$$	\bar q \gamma_\mu (1 \pm \gamma_5) Q = \left[{\alpha_s
(m_Q)\over \alpha_s (\mu)}\right]^{-6/(33-2N_f)} \bar q \gamma_\mu (1 \pm
\gamma_5) h_v^\Q \,\, . \eqno (46)$$
Eq. (45) holds for $v \cdot k \ll m_{c,b}$ which insures that momentum
transfers associated with the light degrees of freedom are small
compared with the heavy quark masses.

The simplest choice for $L$ in eq. (45) is the vacuum.  Then (45)
gives a relation between decay constants$^{12}$
$$	f_B = \left[{\alpha_s (m_b)\over \alpha_s (m_c)}\right]^{-6/25}
\sqrt{{m_D\over  m_B}} f_D \,\, . \eqno (47)$$
There is evidence from $2$-D QCD in the large $N_c$ limit,$^{13}$ QCD sum
rules,$^{13}$ the nonrelativistic constituent quark model and (most
importantly) Lattice Monte Carlo calculations$^{15}$ that the
$\Lambda_{QCD}/m_c$ corrections to eq. (47) are very large.

In principle, eq. (45) together with data on $B\rightarrow L e\bar
\nu_e$ and $D\rightarrow L \bar e\nu_e$ can be used to determine the
$b\rightarrow u$ weak mixing angle $|V_{ub}|$.  The point is that since
the mixing angles are known for the $D$ decay case experimental data on
$D \rightarrow L\bar e\nu_e$ can be used to determine the r.h.s. of eq.
(45).  There are, however, complications that arise because typically
not all the form factors that characterize these matrix elements are
measurable.  Consider for definitness the care where $L$ is a single
pion.  The $B \rightarrow \pi$ matrix element of the axial current
vanishes (because of parity invariance of the strong interactions) and
the vector current matrix element is characterized by two Lorentz
invariant form factors that can be taken to be functions of $v \cdot k$,
$$	\langle \pi (k) | \bar u \gamma_\mu b| B(v)\rangle =
{1\over\sqrt{2m_B}} [f_+^{(B\rightarrow\pi)} (p_B + k)_\mu +
f_-^{(B\rightarrow\pi)} (p_B - k)_\mu]\,\, , \eqno (48)$$
where $p_B = m_Bv$.  The l.h.s. of eq. (48) is independent of $m_b$
(for large $m_b$) and so
$$	f_+ + f_- \sim {\cal O} (1/\sqrt{m_b})\,\, , \eqno (49a)$$
$$	f_+ - f_- \sim {\cal O} (\sqrt{m_b})\,\, . \eqno (49b)$$
Eq. (48) implies that (for large $m_b$) $f_+ + f_-$ is much smaller than $f_+
- f_-$ or equivalently $f_+ \simeq - f_-$.  Using eq. (45) for $L =
\pi$ gives$^{16}$
$$	(f_+ + f_-)^{(B\rightarrow\pi)} = \left({m_D\over
m_B}\right)^{1/2} \left[{\alpha_s (m_b)\over \alpha_s
(m_c)}\right]^{-6/25} (f_+ + f_-)^{(D\rightarrow\pi)} \,\, , \eqno
(50a)$$
$$	(f_+ - f_-)^{(B\rightarrow\pi)} = \left({m_B\over
m_D}\right)^{1/2} \left[{\alpha_s (m_b)\over\alpha_s
(m_c)}\right]^{-6/25} (f_+ - f_-)^{(D\rightarrow\pi)}\,\, . \eqno
(50b)$$
Neglecting the electron mass the matrix elements for $B\rightarrow \pi e
\bar \nu_e$ and $D\rightarrow\pi \bar e\nu_e$ depend only on $f_+$.
However, eqs. (50) relate $f_+^{(B\rightarrow\pi)}$ to a linear
combination of $f_+^{(D\rightarrow\pi)}$ and $f_-^{(D\rightarrow\pi)}$.
Fortunately we can use $f_+ \simeq -
f_-$ to get  a relation between the physically measurable quantities
$$	f_+^{(B\rightarrow\pi)} = \left[{m_B\over m_D}\right]^{1/2}
\left[{\alpha_s (m_b)\over \alpha_s (m_c)}\right]^{-6/25}
f_+^{(D\rightarrow\pi)}\,\, . \eqno (51)$$
If SU(3) is used instead of isospin then $f_+^{(D\rightarrow\pi)}$ on
the r.h.s. of eq. (51) can be replaced by $f_+^{(D\rightarrow K)}$
which has already been measured.

Naively eq. (51) is valid as long as $v \cdot k \ll m_{c,b}$ (Quark
model estimates suggest that it holds even for $v \cdot k \sim
m_{c,b}$).  However, there are important corrections that arise for very
small $v \cdot k \sim m_\pi$.  They occur because of pole graphs
involving the $B^*$ (for $B \rightarrow \pi$) and $D^*$ (for $D
\rightarrow \pi$) mesons.  The form of the $B \rightarrow \pi$ matrix
element for very small $v \cdot k$ is determined by chiral perturbation
theory which gives$^{17}$
$$	(f_+ + f_-)^{(B\rightarrow \pi)} = - \left({f_B\over
f_\pi}\right) \left[1 - {g v \cdot k\over (v \cdot k + m_{B^{*}} -
m_B)}\right]\,\, , \eqno (52a)$$
$$	(f_+ - f_-)^{(B\rightarrow \pi)} = {- gf_B m_B\over f_\pi (v
\cdot k + m_{B^{*}} - m_B)}\,\, . \eqno (52b)$$
In eqs. (52) $g/f_\pi$ is the $B^*B\pi$ coupling which by heavy quark
flavor symmetry is the same as the $D^* D\pi$ coupling.  It determines
the $D^*$ width
$$	\Gamma (D^{*+} \rightarrow D^0 \pi^+) = {g^2\over 6\pi f_\pi^2}
|\vec k_\pi|^3\,\, , \eqno (53)$$
where $f_\pi \simeq$ 135 MeV.   In the nonrelativistic constituent quark
model $g = 1$ (a similar estimate for the pion nucleon coupling gives
$g_A = 5/3$).

A formula like (52) also holds for $D \rightarrow \pi$ (just replace
all the $B$ subscripts by $D$ subscripts).  Because the $D^* - D$ mass
difference is comparable with $m_\pi$, for $v \cdot k \sim m_\pi$, it is
not a good approximation to neglect $m_{D^{*}} - m_D$ compared with $v
\cdot k$.  Hence, eq. (51) (which neglects order $\Lambda_{QCD}/m_c$
effects like the $D^* -D$ mass difference) doesn't hold in this kinematic
regime.

Our discussion of heavy-hadron to light-hadron semileptonic transitions
has focussed on those aspects that may be useful for extracting
$|V_{ub}|$.  However, it is worth recalling that there are interesting
applications of the heavy quark spin symmetry to such transitions that
don't bear on this issue.  For example, charm quark spin symmetry
constrains the form factors for $\Lambda_c \rightarrow \Lambda \bar
e\nu_e$.$^{18}$
This prediction of charm quark spin symmetry has recently been verified
experimentally.

There are very important applications of heavy quark symmetry for
heavy-hadron to heavy-hadron semileptonic transitions.  The classic
example is $B \rightarrow D e\bar\nu_e$ and $B \rightarrow D^*
e\bar\nu_e$ semileptonic decays.  Lorentz and parity invariance of QCD
imply that (recall heavy meson states are normalized to unity instead of
twice their mass)
$$\langle D(v') | \bar c \gamma_\mu b|B(v)\rangle =
{1\over 2} [\tilde{f}_+ (v + v')_\mu + \tilde{f}_- (v - v')_\mu]\,\, ,
\eqno (54a)$$
$$\langle D^*(v',\epsilon)| \bar c \gamma_\mu \gamma_5 b  |B(v)\rangle =
{1\over 2} [\tilde{f} \epsilon_\mu^* + \tilde{a}_+ (\epsilon^* \cdot v)
(v + v')_\mu $$
$$ ~~~~~~~~~~~~~~~~~~~~~~~~~~~~~~~~~~~~~~~~~~~~~~~~~~~~~~~~~~~~~~~~+
\tilde{a}_- (\epsilon^* \cdot v) (v - v')_\mu]\,\, ,\eqno (54b)$$
$$\langle D^* (v',\epsilon) |\bar c \gamma_\mu b |B(v)\rangle = {1\over
2} i \tilde{g} \epsilon_{\mu\nu\lambda\sigma} \epsilon^{*\nu}
v^{\prime\lambda} v^\sigma \,\, . \eqno (54c)$$
In eqs. (54) the Lorentz invariant form
factors $\tilde{f}_\pm,\tilde{f}, \tilde{a_\pm}$ and $\tilde{g}$ are functions
of $v \cdot
v'$.  Heavy quark spin symmetry implies that all of these form factors
are expressed in terms of a single universal function of $v \cdot v'$
(the Isgur-Wise function) $\xi(v \cdot v'$).  The relationship
is$^{1}$
$$	\tilde{f}_+ = \left[{\alpha_s (m_b)\over \alpha_s
(m_c)}\right]^{-6/25} \xi (v \cdot v') \,\, , \quad \tilde{f}_- = 0 \,\,
\eqno (55a)$$
$$	\tilde{f} = \left[{\alpha_s (m_b)\over \alpha_s
(m_c)}\right]^{-6/25} (1 + v \cdot v') \xi (v \cdot v') \,\, \eqno
(55b)$$
$$	(\tilde{a}_+ - \tilde{a}_-) = - \left[{\alpha_s (m_b)\over
\alpha_s (m_c)}\right]^{-6/25} \xi (v \cdot v') \,\, , \quad (\tilde{a}_+
+ \tilde{a}_-) = 0 \,\, , \eqno (55c)$$
$$	\tilde{g} = \left[{\alpha_s (m_b)\over \alpha_s
(m_c)}\right]^{-6/25} \xi (v \cdot v') \,\, . \eqno (55d)$$
Furthermore the normalization of $\xi$ at the zero recoil point, $v
\cdot v' = 1$, is determined by heavy quark flavor symmetry to
be$^{1,19}$
$$	\xi (1) = 1 \,\, . \eqno (56)$$
The perturbative corrections to eqs. (55) of order $\alpha_s (m_b)$
and $\alpha_s (m_c)$ are calculable and don't cause any loss of
predictive power.$^{20,21}$  At order $\Lambda_{QCD}/m_{c,b}$ new
universal functions enter in the form factors and the predictive power
of heavy quark symmetry is greatly diminished.  However, at zero recoil,
$v \cdot v' = 1$, it has been shown that there are no
$\Lambda_{QCD}/m_{c,b}$ corrections.$^{22}$  This remarkable result is usually
called Luke's theorem and it means that $B \rightarrow D^* e\bar\nu_e$
decay may eventually provide a very accurate determination of
$|V_{cb}|$.  Chiral perturbation theory has been used to analyze the
order $(\Lambda_{QCD}/m_c)^p, p=2,3,...$ corrections at zero recoil.  The
corrections, of this order,  that have a nonanalytic dependence on the pion
mass of the form $\ell n m_\pi$ for $p = 2$ and $(1/m_\pi)^{p-2}$ for $p =
3,4,...$ are calculable.  They are about a few percent in
magnitude.$^{23}$  The divergence that occurs as $m_\pi \rightarrow 0$
makes all these $(\Lambda_{QCD}/m_c)^p, p = 2,3...$ corrections of
comparable importance.

Perhaps the most elegant application of heavy quark symmetry is to the
decay $\Lambda_b \rightarrow \Lambda_c e\bar\nu_e$.   In this situation,
the physics is particularly simple since the $\Lambda_{b,c}$ have
$s_\ell = 0$.  In the $m_{b,c} \rightarrow \infty$ limit all the form
factors are again expressible in terms of a single universal function.
Also, perturbative order $\alpha_s (m_b)$ and $\alpha_s (m_c)$
corrections are calculable
and don't cause any loss of predictive power.  However, in this
case  even the $\Lambda_{QCD}/m_{c,b}$ nonperturbative predictions cause
little loss of predictive power.$^{24}$  They are calculable in terms of the
single quantity $\bar\Lambda (\Lambda_{b,c})$ introduced in Section 4.
Unlike the meson case no new unknown functions of $v  \cdot v'$ arise at
order $\Lambda_{QCD}/m_{c,b}.$

\noindent{\bf 6.  Inclusive Semileptonic Decays}
\smallskip

The inclusive lepton spectrum from semileptonic $B$ decays has undergone
intensive experimental and theoretical study.  Recently there has  been
considerable progress in understanding inclusive semileptonic decay.
Inclusive semileptonic B-decay can be treated in a fashion similar to
deep inelastic scattering.  Using a two step process that consists first
of an operator product expansion and then a transition to the heavy
quark effective theory it can be  shown that $d\Gamma/dq^2 dE_e~ (q^2 =
(p_e - p_{\bar\nu_{e}})^2)$, when suitably averaged over $E_e$, is
calculable.$^{25}$  The leading order result is
$$	{d\Gamma^{(0)}\over d\hat q^2 dy} = \sum_j {|V_{jb}|^2 G_F^2
m_b^5\over 192\pi^3} \theta (1 + \hat q^2 - \rho - {\hat q^2\over y} -y)$$
$$	\cdot \{12 (y - \hat q^2) ( 1 + \hat q^2 - \rho - y) \} \,\, ,
\eqno (57)$$
where
$$	\hat q^2 = {q^2\over m_b^2} ~, \quad \rho = {m_j^2\over m_b^2}~,
\quad y = {2E_e\over m_b} \,\, , \eqno (58)$$
and $j = u$ or $c$.  This agrees with free $b$-quark decay.

The $b$-quark and $c$-quark masses that appear in eq. (57) have a
precise meaning.  These masses are the same heavy quark masses as appear
in eq. (10) which describes the transition from QCD to the heavy quark
effective theory (i.e., pole masses in the heavy quark propagators).

The full power of this method for treating inclusive semileptonic decay
becomes apparent when corrections to the
leading result are considered.  These corrections are of two types,
perturbative $\alpha_s (m_b)$ corrections and nonperturbative
corrections suppressed by powers of $\Lambda_{QCD}/m_b$.  In fact there
are no nonperturbative corrections of order $\Lambda_{QCD}/m_b$.  They
first arise at order $(\Lambda_{QCD}/m_b)^2$.

The nonperturbative corrections to eq. (57) of order
$(\Lambda_{QCD}/m_b)^2$ are proportional to the quantities $G_b(B)$ and
$K_b(B)$ (which were defined in Section 4) and have recently been
calculated.$^{26,27,28}$  They add the following terms to the differential
decay rate$^{27}$
$$	{d^2 \Gamma^{(2)}\over dy d\hat q^2} = \sum_j {|V_{jb}|^2 G_F^2
m_b^5\over 192\pi^3} \Bigg\{\theta (1 + \hat q^2 - \rho - {\hat q^2\over
y} - y)$$
$$	\cdot [12 E_b(B)(2\hat q^4 - 2\hat q^2 \rho + y - 2 \hat q^2 y + \rho
y) + 8 K_b (B) (2 \hat q^2 - \hat q^4 + \hat q^2 \rho - 3y)$$
$$	+ 8 G_b (B) (-\hat q^2 + 2 \hat q^4 - 2 \hat q^2 \rho - 2y -
2\hat q^2 y + \rho y)]$$
$$	+ \delta (1 + \hat q^2 - \rho - {\hat q^2\over y} - y) {1\over y^2}
[12 E_b (B) \hat q^2(y - \hat q^2) (-\hat q^2 + 2y - y^2)$$
$$	+ 4 K_b (B) (-\hat q^6 + 9 \hat q^4 y - 6 \hat q^2 y^2 - 2 \hat
q^4 y^2 - \hat q^2 y^4 + y^5)$$
$$	+ 8 \hat q^2 G_b (B) (y - \hat q^2) (- \hat q^2 + y + y^2)]$$
$$	+ \delta' (1 + \hat q^2 - \rho - {\hat q^2\over y} - y) K_b (B)
{4 \hat q^2\over y^3} (y^2 - \hat q^2)^2 (y  - \hat q^2)\Bigg\}\,\, .
\eqno (59)$$
In eq. (59) $E_b (B) = K_b(B) + G_b (B)$.  The terms in eq. (59)
proportional to $E_b(B)$ and $K_b(B)$ have a simple physical
interpretation.  They can be thought of as arising from a shift in the
$b$-quarks mass and four velocity due to the bound state.

The corrections in eq. (59) are singular along the boundary of the
Dalitz plot, $1 + \hat q^2 - \rho - \hat q^2/y - y = 0$ because of the
delta function and derivative of a delta function.  This singular behavior
is an indication that near the boundary of the Dalitz plot the
theoretical prediction must be smeared over a range of electron energies
to be physically meaningful.  (Perturbative corrections are also
singular here and must also be smeared.)  The range of electron energies must
be large enough that $d^2 \Gamma^{(2)}/dy d \hat q^2$ (when smeared) can be
treated as a small correction to $d^2 \Gamma^{(0)}/dy d\hat q^2$.  This
comparison indicates that near the boundary of the Dalitz plot the
electron energy must be smeared over a range of electron energies
$\Delta E_e \>$ 500 MeV.  The endpoint region of the electron spectrum,
$m_B/2 > E_e > (m_B^2 - m_D^2)/2m_B$, is very important
since $b \rightarrow c$ transitions cannot contribute there.  A
theoretical prediction for the normalization of the electron spectrum in
this region would allow the extraction of $|V_{ub}|$ from experimental
data on the endpoint region electron spectrum.  Unfortunately  the
smearing makes a theoretical prediction (from QCD)  of this normalization
impossible.  Extractions of $|V_{ub}|$ with this method are model
dependent.  Fortunately as discussed in Section 5 there is hope that
$|V_{ub}|$ can be determined from exclusive weak decays.  However, much
further work is needed on estimating the size of the
$\Lambda_{QCD}/m_{c,b}$
corrections in order to assess the viability of this method.$^{29}$  The
method for calculating nonperturbative corrections to semileptonic decay
rates outlined above has also be applied to the inclusive rare decays $B
\rightarrow
X_s e^+ e^-$ and $B\rightarrow X_s \gamma$.$^{30}$

\noindent{\bf 7.  Fragmentation to Quarkonium}

Quarkonium production in high energy processes is dominated by
fragmentation of heavy quarks and gluons.  For example, in $Z^0$ decay
the short distance process $Z^0 \rightarrow \psi {gg}$ is suppressed
relative to the fragmentation process $Z^0 \rightarrow \psi c \bar c$ by
a factor of order $m_c^2/m_Z^2$.  Recently it has been realized that the
process in dependent fragmentation functions for quarkonium production
in high energy experiments are computable.$^{31,32}$

For definitness consider the fragmentation of charm quarks to $\psi$'s.
The fragmentation function to transversely aligned $\psi$'s is$^{33}$
$$	D_{c \rightarrow \psi}^T (z) = {16\over 81m_c^2} \alpha_s
(2m_c)^2 f_\psi^2 {2\over 3} {z (1-z)^2\over (2 - z)^6}$$
$$	\cdot \{ 16 - 32z + 76z^2 - 36z^3 + 6z^4\} \,\, ,\eqno (60)$$
and the total fragmentation function to $\psi$'s (i.e., sum of
longitudinal and transverse polarizations) is$^{30}$
$$	D_{c \rightarrow \psi}^{T+L} (z) = {16\over 81m_c^2} \alpha_s
(2m_c)^2 f_\psi^2 {z(1-z)^2\over (2 - z)^6}$$
$$	\cdot\{16 - 32z + 72z^2 - 32z^3 + 5z^4\}\,\, . \eqno (61)$$
In eqs. (60) and (61) the fragmentation functions are evaluated at a
subtraction point $\mu \approx 2m_c$.  They can be evolved to higher
energies using the Altarelli Parisi equations.  This evolution will induce
a fragmentation  function $D_{g \rightarrow \psi} (z)$.  (At $\mu =
2m_c~ D_{g \rightarrow \psi} (z)$ is of order $\alpha_s (2m_c)^3$.)  In eqs.
(60) and (61) $f_\psi$ is the $\psi$ decay constant, defined by
$$	\langle 0|\bar c \gamma_\mu c| \psi (p, \epsilon)\rangle =
f_\psi m_\psi \epsilon_\mu\,\, . \eqno (62)$$
The leptonic width of the $\psi$ determines that $f_\psi \simeq 410$ MeV.  The
total fragmentation probability $P_{c\rightarrow \psi} = \int
dz D_{c \rightarrow \psi}^{T + L} (z)$ is subtraction point
independent and eq. (61) gives $P_{c \rightarrow \psi} \simeq 2 \times
10^{-4}$.

If the expression in the brace brackets of eqs. (60) and (61) were
identical the $\psi$'s produced by fragmentation are unaligned.
Comparison of eqs. (60) and (61) indicates a very slight preference
for transversely aligned psi's.  Let $\zeta$ be the ratio of transverse to
total fragmentation probabilities
$$	\zeta = {\int^1_0 dz D_{c \rightarrow \psi}^T (z)\over \int^1_0
dz D_{c \rightarrow \psi}^{T+L} (z)}\,\, . \eqno (63)$$
This ratio is $\mu$ independent and using eqs. (60) and (61) we find
$\zeta = 0.69$ to be compared with $\zeta = 2/3$ for the production of
unaligned $\psi$'s.  The ratio $\zeta$ is measurable through the angular
distribution of the leptons in the decay $\psi \rightarrow \ell^+
\ell^-$.  Defining $\theta$ to be the angle between the alignment axis
and the lepton momentum the angular distribution $d\Gamma/d \cos \theta$
in the $\psi$ rest frame has the form
$$	{d\Gamma (\psi \rightarrow \ell^+ \ell^-)\over d \cos \theta}
\,\, \propto \,\, (1 + \alpha \cos^2 \theta) \eqno (64)$$
where
$$	\alpha = {3\zeta - 2\over 2 - \zeta} \,\, , \eqno (65)$$
$\zeta = 0.69$ corresponds to the small asymmetry $\alpha = 0.053$.
There are of course other sources of $\psi$'s that must be taken into
account at this level.  For example fragmentation to $P$-wave quarkonium
that subsequently decays to a $\psi$ can disturb  this prediction for
$\alpha$.   ($\psi$'s can also arise from production of radially excited
states which subsequently decay to $\psi\pi\pi$.  However, this
production mechanism does not correct the value of $\zeta$).

It is interesting to compare the alignment of $\psi$'s produced by
fragmentation with the alignment of $\psi$'s produced by nonleptonic
$B$-decay.  Assuming the four-quark amplitude factorizes a
straightforward calculation of $b \rightarrow \psi s \rightarrow \ell^+
\ell^- s$ gives
$$	\alpha = - {m_b^2 - m_\psi^2\over m_b^2 + m_\psi^2} \approx -
0.46 \,\, , \eqno (66)$$
a value very different from $\psi$'s produced by fragmentation.
Applications of the ideas reviewed in this section to fragmentation to
$B_c$ mesons$^{31,32}$ and baryons containing two heavy quarks$^{34}$
have also been considered in the recent literature.

\noindent{\bf 8.  References}
\smallskip

\item{1.}  N. Isgur and M.B. Wise, Phys. Lett., {\bf B232} 113 (1989); Phys.
Lett., {\bf B237} 527 (1990).

\item{2.}  E. Eichten and B. Hill, Phys. Lett., {\bf B234} 511 (1990); H.
Georgi, Phys. Lett., {\bf B240} 447 (1990).

\item{3.}  N. Isgur and M.B. Wise, Phys. Rev. Lett., {\bf 66} 1130 (1991).

\item{4.}  A.F. Falk and M.E. Peskin, SLAC-PUB-6311 (1993) unpublished.

\item{5.}  M. Lu, M.B. Wise and N. Isgur, Phys. Rev., {\bf D45} 1553 (1992).

\item{6.}  R.L. Jaffe and L. Randall, CTP\#2189 (1993) unpublished.

\item{7.}  A.F. Falk, B. Grinstein and M.E. Luke, Nucl. Phys., {\bf B357}
185 (1991); E. Eichten and B. Hill, Phys. Lett., {\bf B243} 427 (1990).

\item{8.}  G.P. Lepage and B.A. Thacker, Nucl. Phys. B (Proc. Suppl.), {\bf
4} 199 (1988).

\item{9.}  M. Luke and A.V. Manohar, Phys. Lett., {\bf B286} 348 (1992); M.J.
Dugan, M. Golden and B. Grinstein, Phys. Lett., {\bf B282} 142 (1992).

\item{10.}  U. Aglietti, Phys. Lett., {\bf B281} 341 (1992).

\item{11.}  Z. Guralnik and A. Manohar, Phys. Lett., {\bf B302} 103 (1993).

\item{12.}  M.B. Voloshin and M.A. Shifman, Sov. J. Nucl. Phys., {\bf 45}
292 (1987); H.D. Politzer and M.B. Wise, Phys. Lett., {\bf B206} 681
(1988).

\item{13.}  M. Burkhardt, Phys. Rev., {\bf D46} 1924 (1992); Phys. Rev. D
{\bf 46} 2751 (1992); M. Burkhardt and E.S. Swanson, Phys. Rev., {\bf D46}
5083 (1992); B. Grinstein and P.F. Mende, Phys. Rev. Lett., {\bf 69}
1018 (1992).

\item{14.}  M. Neubert, Phys. Rev., {\bf D45} 2451 (1992).

\item{15.}  See talk by P. Mackenzie and references therein.

\item{16.}  N. Isgur and M.B. Wise, Phys. Rev., {\bf D42} 2388 (1990).

\item{17.}  M.B. Wise, Phys. Rev., {\bf D45} 2188 (1992); G. Burdman and J.
Donoghue, Phys. Lett., {\bf B208} 287 (1992); T.M. Yan, H.Y. Cheng, C.Y.
Cheng, G.L. Lin, Y.C. Lin and H.L. Yu, Phys. Rev., {\bf D46} 1148
(1992).

\item{18.}  T. Mannel, W. Roberts and Z. Ryzak, Nucl. Phys., {\bf B355}
38 (1991); F. Hussain, J.H. Korner, M. Kramer and G. Thompson, Z. Phys.,
{\bf C51} 321 (1991).

\item{19.}  S. Nussinov and W. Wetzel, Phys. Rev., {\bf D36} 130 (1987); M.A.
Shifman and M.B. Voloshin, Sov. J. Nucl. Phys., {\bf 47} 511 (1988).

\item{20.}  A.F. Falk, H. Georgi, B. Grinstein and M. Wise, Nucl. Phys., {\bf
B343} 1 (1990); A. Falk and B. Grinstein, Phys. Lett., {\bf B247} 406
(1990); Phys. Lett., {\bf B249} 314 (1990).

\item{21.}  M. Neubert, Phys. Rev., {\bf D46} 2212 (1992).

\item{22.}  M.E. Luke, Phys. Lett., {\bf B252} 447 (1990).

\item{23.}  L. Randall and M.B. Wise, Phys. Lett., {\bf B303} 139 (1993).

\item{24.}  H. Georgi, B. Grinstein and M.B. Wise, Phys. Lett., {\bf B252}
456 (1990).

\item{25.}  J. Chay, H. Georgi and B. Grinstein, Phys. Lett., {\bf B247}
399 (1990).

\item{26.}  I.I. Bigi, M. Shifman, N.G. Uraltsev and A.I. Vainshtein, Phys.
Rev. Lett., {\bf 71} 496 (1993).

\item{27.}  B. Blok, L. Koyrakh, M. Shifman and A.I. Vainshtein,
NSF-ITP-93-68, (1993) unpublished;  A.V. Manohar and M.B. Wise,
CALT-68-1883 (1993) unpublished.

\item{28.}  T. Mannel, IKDA-93/26 (1993) unpublished.

\item{29.}  C.O. Dib and F. Vera, Phys. Rev., {\bf D47} 3938 (1993).

\item{30.}  A.F. Falk, M. Luke and M. Savage, SLAC-PUB-6317 (1993)
unpublished.

\item{31.}  E. Braaten, K. Cheung and T.C. Yuan, NUHEP-TH-93-2 (1993)
unpublished; E. Braaten and T.C. Yuan, NUHEP-92-23 (1993) unpublished;
E. Braaten, K. Cheung and T.C. Yuan, NUHEP-TH-93-6 (1993) unpublished.

\item{32.}  Y.Q. Chen and Y.P. Kuang, Phys. Rev., {\bf D46} 1165 (1992);
Y.Q. Chen, PhD Thesis, The Institute of Theoretical Physics, Academia
Sinica, P.R. of China (1992).

\item{33.}  A.F. Falk, M. Luke, M. Savage and M.B. Wise, Phys. Lett.,
{\bf B312} 486 (1993).

\item{34.}  A.F. Falk, M. Luke, M. Savage and M.B. Wise, CALT-68-1868
(1993) unpublished.
\end